\documentclass[pre,article]{revtex4}
\usepackage{graphicx}
\usepackage{amsmath}
\usepackage{amsfonts}
\usepackage{amssymb}
\providecommand{\U}[1]{\protect\rule{.1in}{.1in}}

\begin{document}

\title{\ A statistical physics of stationary and metastable states: \\description of
the plasma column experimental data}
\author{A. Cabo$^{1}$, S. Curilef$^{2}$, A. Gonz\'alez$^{1}$, N. G.
Cabo-Bizet$^{3,5}$ and C. A. Vera$^{4}$ }

\affiliation{$^{1}$\textit{Theoretical Physics Department,
Instituto de Cibern\'{e}tica, Matem\'{a}tica y F\'{\i}sica, Calle
E, No. 309, Vedado, La Habana, Cuba. }\\
$^{2}$\textit{Departamento de F\'{\i}sica, Universidad Cat\'{o}lica del Norte, Av. Angamos 0610, Antofagasta,  Chile.}\\
$^{3}$\textit{Departamento de F\'{\i}sica,  Centro
de Aplicaciones Tecnol\'{o}gicas y Desarrollo Nuclear (CEADEN),
Calle 30, esq. a 5ta Ave, Miramar, La Habana, Cuba. }\\
$^{4}$\textit{Instituto de F\'{\i}sica,
Universidad de Antioquia, Medellin, Colombia.}\\
$^{5}$\textit{Physikaliches Institut der Universit\"at Bonn.
Nussallee 12, Bonn, Deutschland.}}

\begin{abstract}
\noindent We propose a statistical mechanics for a general class
of stationary and metastable equilibrium states. For this purpose,
the Gibbs extremal conditions are slightly modified in order to be
applied to a wide class of non-equilibrium states. As usual, it is
assumed that the system maximizes the entropy functional $S$,
subjected to the standard conditions; \emph{i.e.,} constant energy
and normalization of the probability distribution. However, an
extra conserved constraint function $F$ is also assumed to exist,
which forces the system to remain in the metastable configuration.
\ Further, after assuming additivity for two quasi-independent
subsystems, and that the new constraint commutes with density
matrix $\rho$, it is argued that $F$ should be an homogeneous
function of the density matrix, at least for systems in which the
spectrum is sufficiently dense to be considered as continuous. The
explicit form of $F$ turns to be $F(p_{i})=p_{i}^{q}$, where $p_i$
are the eigenvalues of the density matrix and $q$ is a real number
to be determined. This $q$ number  appears as a kind of Tsallis
parameter  having the interpretation of the order of homogeneity
of the constraint $F$.  The procedure is applied to describe the
results of the plasma experiment of Huang and Driscoll. The
experimentally measured density is predicted with a similar
precision  as it is done with the use of the extremum of the
enstrophy and Tsallis procedures. However, the present results
define the density at all the radial positions. In particular, the
smooth tail shown by the experimental distribution turns to be
predicted  by the procedure. In this way, the scheme avoids the
non-analyticity of the density profile at large distances arising
in both of the mentioned alternative procedures.

\bigskip

\bigskip

\end{abstract}


 \maketitle

\section{Introduction}
Nowadays, a relevant topic of research in statistical physics is the
characterization of metastable and non-equilibrium states of
physical systems \cite{degroot,basic2,basic3, basic4,tsallis}.
Multiple efforts have been made in order to explain this kind of
behavior, but the discussion intensively continues because of the
large variety and complexity of physical processes involved
\cite{tsallis,driscol,boghosian,smith}.

In this work we intend to present  a more complete discussion of
the statistical description previously sketched in Ref.
[\onlinecite{curilef}]. We expect to describe the statistical
physics of a large class of metastable and stationary states. Our
main suggestion to deal with these situations is that statistical
properties of the metastable or stationary non-equilibrium states
depart from the Gibbs thermal states in equilibrium in the
following sense: \emph{the metastable or non-equilibrium
configuration may  be the results of an additional dynamical
constraint, maintaining the system for a while in the metastable
state, if the restriction  is only an approximate one; and
retaining the system in a time independent configuration,  in the
case of non-equilibrium stationary states. After the removal of
the condition, the system is allowed to evolve in a thermal state
described by the Bloch density matrix}.  Hence, the time
independence of the statistical properties of the systems leads to
commutation between the   density matrix and the Hamiltonian
during the time for which the metastable or stationary state
exists. As it was mentioned,  we will assume that an extra
constraint $F$ is conserved in time, then $F$ will also commute
with the Hamiltonian. In the large class of systems having a
non-degenerate spectrum of the energy, the Hamiltonian $H$, the
density matrix $\rho$ and the extra constraint $F$ can be
diagonalized in the same common basis of eigenstates. The same
property is not necessarily valid when the energy spectrum is non
degenerated. From now on, we will simply restrict the discussion
to systems exhibiting the mentioned commutativity among $H$,
$\rho$ and $F$.

Let us assume that the extra constraint defines the metastable
state. Then, it is possible to note the existence of some related
conserved constraints. A particular interesting one is given by the
product operator $HF$. Note, that the assumed commutativity of $\rho
$ with $F$ implies the possibility of expressing $F$ as a function
of $\rho.$   The above remark, motivated in us  the idea of
constructing  a special analytical form of the condition to be
incorporated in the Lagrange multiplier scheme for  the maximization
of the entropy. This condition will be constructed as a modified
expectation value of the energy, in which in the density matrix in
the usual mean value is substituted by the constraint function $F$.
This quantity results  to be conserved as a direct consequence of
the validity of the constraint $F$.

 As a result of the above construction, it follows that, when the
additivity condition of the resulting statistical description is
assumed to be valid for two approximately independent subsystems,
the constraint function $F$ \ should have the Tsallis homogeneous
structure $F(\rho)=C_{q}\rho^{q}$ where $q$ is a real number to be
determined. It should be underlined that this conclusion follows
for systems in which the energy spectrum  is sufficiently dense
for to be considered as continuous. For such systems, the
analysis, seems to indicate an interpretation of the Tsallis $q$
parameter as corresponding to the degree of homogeneity of the
constraint $F$  after to be represented as a function of the
density matrix $\rho$.

In order to start investigating its implications, the procedure
was applied in this work to describe the experimental data
obtained from  the plasma experiment performed by \ Huang and
Driscoll \cite{driscol}. The measurements of  the electron density
 were chosen as an input of an iterative procedure  for
solving the equations for the probability density $\rho.$ The
results predicted for the densities at radial distances at which the
their values are not small, describe the experimental results with
similar quality as the ones following from  two theoretical schemes
existing for this problem. They are the minimization of the
enstrophy on one case and the maximization of the Tsallis entropy in
the other one \cite{driscol, boghosian}. It is known that these
procedures are equivalent and their applications to the considered
problem are recognized as  main studies existing in recent
literature \cite{boghosian}. In connection with the zone of small
radial distances,  both of the analysis  describe this region with
similar precision and their results are close to the experimental
data. However, at large radial distances, where the density is small
enough, their results strongly deviate from the experimental
measurements reported by Huang and Driscoll \cite{driscol}. Our
results radically improve the outcome of the aforementioned
theoretical procedures in this large distance region. In this sense,
the presented approach gives a new description of the smooth
vanishing tail of the density profile, which has been experimentally
measured. Each one of the alternative procedures needs to justify
the presence of a non-analytical behavior of the density
distributions in the region of the tail of small densities, as
coming form the requirement of the non-negative character of the
electron density. In the here proposed discussion, such a procedure
is not becoming necessary and, as noted before, the dependence of
the density as a function of the radial distance is predicted to
smoothly decay.

  Our procedure has been applied to the quasiequilibrium state of
excitonic polaritons \cite{carlos}. These polaritons are
quasibosonic quasiparticles arising from the strong interaction of
excitons and confined light modes in a microcavity. A finite model
is studied in Ref. [\onlinecite{carlos}] with a help of a master
equation for the density matrix. The result is that the density
matrix approximately commutes wit the Hamiltonian.  We assumed
that the quasiequilibrium is the result  of a constraint in phase
space  and showed that an analytical expression for the density
matrix, resulting from our procedure  gives a reasonable fit to
the numerical results.

 The paper proceeds as follows. In Section II  we
present the basic elements of the proposed statistical
description.  Section \ref{parameterq}, then continues by arguing
that the assumption of additivity in the statistical description
implies that the constraint function should be a simple $q$ power
of the density matrix. In Section \ref{plasma} the procedure is
applied to describe the results of the plasma experiment of Huang
and Driscoll. Finally, Section \ref{summary} resume the results of
the work.

\bigskip

\section{A  statistical mechanics for a class of metastable
and stationary states }

Let us consider a physical system having a quantum dynamics
described by a Hamiltonian $H$. In the Gibbs approach, the
properties of the system in thermal equilibrium are contained in
the Bloch density matrix
\[
\rho=\exp(-\frac{H}{kT}),
\]
which satisfies $[H,\rho]=0$ \
and determines the conditional maximum of the entropy functional \ %
\begin{align*}
S  &  =-Tr[\rho\log(\rho)]+\alpha(Tr[\rho H]-E)+\\
&  +\beta(Tr[\rho]-1),
\end{align*}
where $\alpha$ and $\beta$ are Lagrange multipliers corresponding
to imposing two conditions: the conservation of the energy $E$ and
the normalization of the trace of the density operator $\rho$. The
latter condition is placed in order to furnish a probability
interpretation to the diagonal elements of $\rho$, this is
$Tr[\rho]=1$. Eventually, additional conserved quantities can be
added by  also introducing their corresponding multipliers.

Our main assumption in this work is that during a large relaxation
time $\tau$, the considered physical systems are forbidden to
reach the thermal equilibrium state, as a consequence  of the
action of extra  constraints. These restrictions are assumed to be
dynamically generated and able to obstructs during macroscopic
time intervals the standard evolution which normally drops the
density matrix to the Bloch form. In other words, we propose the
existence of a conserved quantity $F$, whose effects over the
motion of the system is to delay the usual evolution, then leading
the system to an intermediate equilibrium state  differing from
the Gibbs thermal one: the metastable or non equilibrium
stationary state. Therefore, the system is compelled to remain in
those states during the time lapse in  that the extra constraint
works. Thus, as a result of the precedent assumptions we will
have:
\[
\lbrack H,F]=0.
\]
Therefore, the constraint $F$ and the density matrix $\rho$ both
commute with the Hamiltonian. As mentioned before, in the large
class of systems in which $H$ has a non degenerated spectrum,  it
is possible to simultaneously diagonalize $H,$ $\rho$ and $F$,
within a common basis of eigenfunctions. However, in a more
general situation, we will simply restrict the discussion to the
cases in that all the three quantities commute among them. Thus,
the conserved constraint $F$ can be expressed as a certain
function of the density matrix  $F=F(\rho)$. From the mentioned
assumptions if follows that the quantity  $F(\rho)H $  is also
conserved
\begin{align*}
\lbrack F(\rho)H,H] &  =0.
\end{align*}
Further,  the evolution of the system in the considered metastable
or stationary states, also implies the  time invariance of the
specially constructed expectation value \ \
\[
\frac{Tr[F(\rho)H]}{Tr[F(\rho)]}.
\]
 Now, let us precisely state our main dynamical principle: the metastable
 and stationary states are determined by the maximization of  entropy
$S$ subjected to the conditions of conservation of normalization,
energy and the aforementioned expectation value. Then, the usual
constraints plus the extra one,  after multiplied by their
corresponding Lagrange multipliers, are added to the entropy
functional to construct the modified  form of the Gibbs Lagrange
multiplier scheme. The functional takes the explicit form%
\begin{align*}
S  &  =-Tr[\rho\log(\rho)]+\alpha(Tr[\rho H]-E)+\\
&  +\beta(Tr[\rho]-1)+\gamma(\frac{Tr[F(\rho)H]}{Tr[F(\rho)]}-E_{C})\\
&  =-\sum_{i}p_{i}\log(p_{i})+\alpha(\sum_{i}p_{i}\epsilon_{i}-E)+\beta
(\sum_{i}p_{i}-1)+\\
&
+\gamma(\frac{\sum_{i}F(p_{i})\epsilon_{i}}{\sum_{i}F(p_{i})}-E_{F}),
\end{align*}
where $p_i,  i=1,2,3...$, are the diagonal elements of the density
matrix $\rho= \sum_i p_i |i><i| $.

\
\section{Tsallis $q$ parameter from the additivity of the
description}\label{parameterq}

  Let us now consider implications of the modified statistical
physics being introduced. They  follow after assuming the additivity
in the statistical description for the  combination of two quasi
independent systems. That is, each one of the subsystems will be
assumed to be in the same kind of metastable state, showing
analogous statistical properties as the whole body. Then, let us
consider a pair of such systems that weakly interact between them.
Any one of the bodies is considered in the  same kind of metastable
state. The entropy and the constraint functions of the first system
can be written as
\begin{align*}
S^{(1)} &  =-\sum_{i}p_{i}^{(1)}\log(p_{i}^{(1)})\\
E^{(1)} &  =\sum_{i}p_{i}^{(1)}\epsilon_{i}^{(1)},\text{ \ \ \ }1=\sum
_{i}p_{i}^{(1)},\\
E_{F}^{(1)} &  =\frac{\sum_{i}F(p_{i}^{(1)})\epsilon_{i}^{(1)}}{\sum
_{i}F(p_{i}^{(1)})}.%
\end{align*}

Analogously, the same  quantities for the second system may be
expressed as
\begin{align*}
S^{(2)}  &  =-\sum_{i}p_{i}^{(2)}\log(p_{i}^{(2)})\\
E^{(2)}  &  =\sum_{i}p_{i}^{(2)}\epsilon_{i}^{(2)},\text{ \ \ \ }1=\sum
_{i}p_{i}^{(2)},\\
E_{F}^{(2)}  &
=\frac{\sum_{i}F(p_{i}^{(2)})\epsilon_{i}^{(2)}}{\sum
_{i}F(p_{i}^{(2)})}.%
\end{align*}
Now let us consider that the description is also valid for the
combination of both systems. Then, the entropy and constraints can
be expressed as follows
\begin{align*}
S^{(1,2)}  &  =-\sum_{(i,j)}p_{(i,j)}^{(1,2)}\log(p_{(i,j)}^{(1,2)})\\
E^{(1,2)}  &  =\sum_{(i,j)}p_{(i,j)}^{(1,2)}\epsilon_{(i,j)}^{(1,2)},\text{
\ \ \ }1=\sum_{(i,j)}p_{(i,j)}^{(1,2)},\\
E_{F}^{(1,2)}  &  =\frac{\sum_{(i,j)}F(p_{(i,j)}^{(1,2)})\epsilon
_{(i,j)}^{(1,2)}}{\sum_{(i,j)}F(p_{(i,j)}^{(1,2)})},
\end{align*}
where the pair \ $(i,j)$ indicates  the state of the composite
system in terms of the indices of the states of the  single
systems as $|i,j\rangle=|i\rangle\times|j\rangle$.  The assumed
separability of the two bodies allows to write a relation for the
probability of the combined states $|i,j\rangle$ in the following
way
\begin{align*}
p_{(i,j)}^{(1,2)} &  =p_{i}^{(1)}p_{j}^{(2)}\\
\sum_{(i,j)}p_{i}^{(1)}p_{j}^{(2)} &  =\sum_{i}p_{i}^{(1)}\sum_{j}p_{j}%
^{(2)}\\
&  =1.
\end{align*}
Thus,  the probability constraint for the individual subsystems is
valid for the probability constraint of the composite system.

The additivity of entropies follows in a similar way %
\begin{align*}
S^{(1,2)}  &  =-\sum_{(i,j)}p_{(i,j)}^{(1,2)}\log(p_{(i,j)}^{(1,2)})\\
&  =-\sum_{(i,j)}p_{i}^{(1)}p_{j}^{(2)}\log(p_{i}^{(1)}p_{j}^{(2)})\\
&  =-\sum_{i}p_{i}^{(1)}\log(p_{i}^{(1)})-\sum_{j}p_{j}^{(2)}\log(p_{j}%
^{(2)})\\
&  =S^{(1)}+S^{(2)}.
\end{align*}
We have the same outcome for the addition of energies
\begin{align*}
E^{(1,2)} &  =\sum_{(i,j)}p_{(i,j)}^{(1,2)}\epsilon_{(i,j)}^{(1,2)},\\
&  =\text{ }\sum_{(i,j)}p_{i}^{(1)}p_{j}^{(2)}(\epsilon_{i}^{(1)}+\epsilon
_{j}^{(2)})\\
&  =\sum_{(i,j)}p_{i}^{(1)}p_{j}^{(2)}\epsilon_{i}^{(1)}+\sum_{(i,j)}%
p_{i}^{(1)}p_{j}^{(2)}\epsilon_{j}^{(2)}\\
&  =\sum_{i}p_{i}^{(1)}\epsilon_{i}^{(1)}+\sum_{j}p_{j}^{(2)}\epsilon
_{j}^{(2)}\\
&  =E^{(1)}+E^{(2)}. %
\end{align*}
Let us come to the main new element considered in this section.
Now we  assume that the $q$-expectation value of the energy also
satisfies the statistical independence condition for the combined system as%
\begin{equation}
F(\rho_{(i,j)}^{(1,2)})=F(\rho_{i}^{(1)})F(\rho_{j}^{(2)}). \label{indep}%
\end{equation}
Therefore, the $q$-expectation value also satisfies the additivity
properties, as follows%
\begin{align*}
E_{F}^{(1,2)} &  =\frac{\sum_{(i,j)}F_{(i,j)}^{(1,2)}\epsilon_{(i,j)}^{(1,2)}%
}{\sum_{(i,j)}F_{(i,j)}^{(1,2)}},\\
&  =\text{ }\frac{\sum_{(i,j)}F_{i}^{(1)}F_{j}^{(2)}(\epsilon_{i}%
^{(1)}+\epsilon_{j}^{(2)})}{\sum_{(i,j)}F_{i}^{(1)}F_{j}^{(2)}}\\
&  =\frac{\sum_{(i,j)}F_{i}^{(1)}F_{j}^{(2)}\epsilon_{i}^{(1)}}{\sum
_{(i,j)}F_{i}^{(1)}F_{j}^{(2)}}+\frac{\sum_{(i,j)}F_{i}^{(1)}F_{j}%
^{(2)}\epsilon_{j}^{(2)}}{\sum_{(i,j)}F_{i}^{(1)}F_{j}^{(2)}\epsilon_{i}%
^{(1)}}\\
&  =\frac{\sum_{i}F_{i}^{(1)}\epsilon_{i}^{(1)}}{\sum_{i}F_{i}^{(1)}}%
+\frac{\sum_{j}F_{j}^{(2)}\epsilon_{j}^{(2)}}{\sum_{i}F_{i}^{(1)}}\\
&  =E_{F}^{(1)}+E_{F}^{(2)}.%
\end{align*}
However, this condition also imposes a strong restriction on the
possible forms of the function $F$ defining the newly introduced
expectation value. In order to see this central point of the
presentation, the statistical independence condition (\ref{indep})
will be rewritten. Consider  $F$ as expanded in powers of its
argument  $x$, in the general form
\begin{equation}
F(x)=x^{\nu}\sum_{n=0}^{\infty}f_{n}\text{ }x^{n}.\label{expansion}
\end{equation}
Combining Eqs.(\ref{indep}) and (\ref{expansion}), it is possible to write  %
\begin{widetext}
\[
(p_{i}^{(1)}p_{j}^{(2)})^{\nu}\sum_{n=0}^{\infty}f_{n}(\text{ }p_{i}%
^{(1)})^{n}(\text{ }p_{j}^{(2)})^{n}=(p_{i}^{(1)})^{\nu}(p_{j}^{(2)})^{\nu
}\sum_{n=0}^{\infty}\sum_{m=0}^{\infty}f_{n}f_{m}(\text{ }p_{i}^{(1)}%
)^{n}(\text{ }p_{j}^{(2)})^{m}.%
\]
\end{widetext}
After a suitable rearrangement  of the variables, it follows
\[
0=\sum_{n=0}^{\infty}\sum_{m=0}^{\infty}f_{n}(\text{ }\rho_{i}^{(1)}%
)^{n}(\text{ }\rho_{j}^{(2)})^{m}(\delta_{nm}-f_{m}).
\]
Next, let us assume  that the domain of the probability $p_i$ for
all the values indices $i$ is a continuous set. Then, the
completeness of the basis formed by the  powers of a variable implies  \ %
\[
f_{n}(\delta_{nm}-f_{m})=0,\text{ \ \ for all \ }m\text{ and }n\text{ }.%
\]
At this point, let us suppose that $f_{m}\neq 0,$ for a particular
value of $m=m_o$. Then, the validity of the above relation directly
implies that  $f_m=0$ for all the values of $m\neq m_o$.
 Therefore,  the explicit form of the  operator $F$ as a function of
 $\rho$ can be written in the following simple  way
\begin{eqnarray}
F(\rho)   &= &f_{m_{o}}\rho^{m_{o}+\nu}  \nonumber\\
&=&f_{q}\text{ }\rho^{q} \label{qtsallis}\\
&=&\sum_i p^{q}_i |i><j| \nonumber
\end{eqnarray}
The parameter  $q$ is a real number which value  should  be
dynamically determined.
 The above conclusion  complete the argue indicating  the Tsallis
mean value structure of the modified average  imposing  the
dynamical constraint $F$.
\subsection{Extremal entropy equations}

Let us examine in this subsection the $F$ dependent modified
expectation value. As noted before, its  most interesting property
is the full coincidence with the Tsallis $q$-expectation value of
the energy
\begin{align*}
E_q \equiv
E_F=\frac{\sum_{i}p_{i}^{q}\epsilon_{i}}{\sum_{i}p_{i}^{q}},
\end{align*}
where  the index $q$ is the power of the probability in
(\ref{qtsallis}).  The  modified entropy functional takes the form
\begin{align*}
S  &  =-\sum_{i}p_{i}\log(p_{i})+\alpha(\sum_{i}p_{i}\epsilon_{i}%
-E)+\beta(\sum_{i}p_{i}-1)+\\
&
+\gamma(\frac{\sum_{i}p_{i}^{q}\epsilon_{i}}{\sum_{i}p_{i}^{q}}-E_{q}).
\end{align*}
and the Lagrange extremum equations following from the functional are%
\begin{align*}
\frac{\partial S}{\partial p_{i}}, i=1,2,...   &  =0,\text{ \ }\frac{\partial S}%
{\partial\alpha}=0,\\
\frac{\partial S}{\partial\beta}  &  =0,\text{ \ \ }\frac{\partial S}%
{\partial\gamma}=0.
\end{align*}
 Their explicit calculation  leads to the following set of
coupled equations for the eigenvalues of the density matrix $\rho$%
\begin{align*}
1-\alpha\text{ }\epsilon_{i}-\beta &  =-\log(p_{i})+\gamma\frac{q\text{ }%
p_{i}^{q-1}}{\sum_{i}p_{i}^{q}}(\epsilon_{i}-E_{q}),   i=1,2,...\\
E  &  =\sum_{i}p_{i}\epsilon_{i},\\
1  &  =\sum_{i}p_{i},\\
E_{q}  &  =\frac{\sum_{i}p_{i}^{q}\epsilon_{i}}{\sum_{i}p_{i}^{q}}.
\end{align*}

After some algebra, we obtain the integral relation
\begin{align*}
0  &  =-\sum_{i}p_{i}\log(p_{i})+\gamma\sum_{i}\frac{q\text{ }p_{i}^{q}}%
{\sum_{i}p_{i}^{q}}(\epsilon_{i}-E_{q})-1+\alpha E+\beta,\\
&  =-\sum_{i}p_{i}\log(p_{i})-1+\alpha E+\beta\\
&  =S+\alpha E-1+\beta.
\end{align*}
The last expression  establishes a connection among the entropy,
energy and the Lagrange multipliers. As usual, it seems useful to
be employed in constructing  generalizations of the free energy
and other thermodynamical potentials and relations. However, we
will not deal with this discussion here.\

\section{Metaequilibrium states in electron plasma columns}\label{plasma}

In this section, we apply the statistical procedure being
investigated to an important example: the relaxation of a 2D
turbulence to a metaequilibrium state. A relevant experiment related
with this problem  was performed and discussed by Huang and Driscoll
\cite{driscol}. They compared results from several theoretical
approaches to their measured data. The theoretical scheme which
furnish  the best approximation, is the minimum enstrophy model. The
experiment considered the 2D dynamics of a electron plasma fluid.
The authors were able to identify and measure properties of a
metaequilibrium state (MES) for a plasma sample in which a
turbulence initially occurs within a rotating magnetized cylindrical
electron column. The turbulence relaxes to a long-lasting
metaequilibria state with axial symmetry.  The relevant conserved
quantities of the electron column in the MES are the following:
energy $H$, entropy $S$, angular momentum $P_{\theta}$ and number of
particles $N_L$.

We start from the Poisson equation  in order to relate the potential
$\Psi=\phi\frac{1}{e N_L}$ to the particle density $\rho=n
\frac{R_w^2}{N_L}$, where $\phi$ and $n$ are the potential and the
particle density (in the international system of units). The
parameter $e$ is the electron charge and $R_w$ the radius of the
rotating drum. In agreement with Ref. \cite{driscol} all magnitudes
are expressed in a special set of units, in which the cylinder
radius  is equal to 1. The Poisson equation for $\Psi$ is
\begin{eqnarray}
\bigtriangledown^2 \Psi&=&4 \pi \rho(r)\label{poisson}\\
&=&\frac{1}{r}\frac{d}{d r}\left(r \frac{d \Psi(r)}{d
r}\right).\nonumber
\end{eqnarray}

The solution of this equation can be obtained in the form
\begin{eqnarray}
\Psi(r)&=&\int_{0}^{1}2\pi r'd r' F^S_0(r,r')\rho(r'), \\
F^S_0(r,r')&=&2\ln(r)\Theta(r-r')+2\ln(r')\Theta(r'-r), \nonumber
\end{eqnarray}
where the gauge freedom of the potential has been employed to obtain
a symmetric expression of the Green function $F^S_0(r,r')$. Note
that this fact may introduce a change in the value of the total
electrostatic energy with respect  to the treatment in Ref.
[\onlinecite{driscol}].

 The  entropy $S$, normalization $N$, energy $H$, angular
momentum $P_{\theta}$ and the constraint $E_q$ are given by the
expressions
\begin{eqnarray}
H&=&-\frac{1}{2}\int_{0}^{1}2 \pi r dr \Psi(r) \rho(r), \label{parameters}\\
P_{\theta}&=&\int_{0}^{1}d^2r (1-r^2)\rho(r), \nonumber\\
N&=&1=\int_{0}^{1}2 \pi r dr \rho(r),\label{norma}\\
S&=&-\int_{0}^{1}2 \pi r dr \rho(r)\ln[\rho(r)],\nonumber\\
E_q &=&\frac{\int \rho^q(r)\left(-\frac{\Psi(r)}{2}\right)d^2
r}{\int \rho^q(r)d^2 r}.\nonumber
\end{eqnarray}

We should then  find the extremum of the functional
\begin{eqnarray}
A&=&-\int \rho(r)\ln[\rho(r)]d^2 r +\alpha_E
\left(H-H^0\right)\nonumber\\&+&\alpha_N\left(N-1\right) +\alpha_P
\left(P_{\theta}-P^0_{\theta}\right)\nonumber\\&+&\gamma
\left(E_q-E^0_q \right).\nonumber
\end{eqnarray}
The Lagrange multipliers are given by $\alpha_E, \alpha_P, \alpha_N$
and $\gamma$. The corresponding Euler equations are
\begin{eqnarray}
\ln [ \rho(\omega)]&=&(\alpha_N -1)+ \alpha_P (1-\omega^2) \label{extremo}\\
&-&\alpha_E \int d^2 r F^S_0(r,\omega) \rho(r)
\nonumber \\
&+& \frac{\gamma}{\int d^2 r \rho^q (r) }\left(-\frac{1}{2}q
\rho^{q-1}(\omega)\Psi(\omega)\right) \nonumber
\\&+&\frac{\gamma}{\int d^2 r \rho^q (r) }\left(-\frac{1}{2}\int d^2 r \rho^q(r) F^S_0(r,\omega)\right)\nonumber\\
&+&\frac{\gamma}{\int d^2 r \rho^q (r) }\left(-E_q q
\rho^{q-1}(\omega) \right).\nonumber
\end{eqnarray}
The integrated  version of the above relation is
\begin{eqnarray}
S+(\alpha_N-1)+2\alpha_E H + \gamma E_q+\alpha_P P_{\theta}=0.
\label{integral}
\end{eqnarray}

We start from the measured  data, $\rho^{(0)}$, for the electron
density, given in Ref. \cite{driscol}, and solve iteratively for
$\rho$ in a discrete version  of relation (\ref{extremo})
\begin{eqnarray}
&&\ln [ \rho(\omega_j)]=(\alpha_N -1)+ \alpha_P (1-\omega_i^2) \label{extremo2}\\
&-&\alpha_E \sum_i 2 \pi r_i \Delta r_i F^S_0(r_i,\omega_j)
\rho(r_i)
\nonumber \\
&+& \frac{\gamma}{\sum_i 2 \pi r_k \Delta r_k \rho^q (r_k)
}\left(-\frac{1}{2}q \rho^{q-1}(\omega_j)\Psi(\omega_j)\right)
\nonumber
\\&+&\frac{\gamma}{\sum_i 2 \pi r_k \Delta r_k \rho^q (r_k) }
\times\nonumber\\
&&\left(-\frac{1}{2}\sum_i 2 \pi r_i \Delta r_i \rho^q(r_i)F^S_0(r_i,\omega_j)\right)\nonumber\\
&+&\frac{\gamma}{\sum 2 \pi r_k \Delta r_k \rho^q (r_k) }\left(-E_q
q \rho^{q-1}(\omega_j) \right),\nonumber
\end{eqnarray}
to obtain consecutive approaches to the density distribution by
exponentiating the logarithm in the left hand side (l.h.s.) of the
relation (\ref{extremo2}). Now, the indices i,j=1,2,... indicate the
increasing values of all the radial positions $r_i$ (or $w_i$) at
which the electron density measurements were reported in Ref.
[\onlinecite{driscol}]. The initial density function
$\rho^0(\omega_j)$ was constructed by assigning the density value
measured  at each point $w_j$ in Ref. [\onlinecite{driscol}]. The
distance elements $\Delta r_k $ are defined as the differences
$\Delta r_k=\frac{r_{k+1}-r_{k-1}}{2}, k=2...N-1$. But $\Delta
r_1=r_2-r_1$ and $\Delta r_N=r_{N}-r_{N-1}$.
 Let us denote the l.h.s. of
Eq.(\ref{extremo2}) evaluated in the density values obtained in Ref.
[\onlinecite{driscol}] as $L^0_j$,  and the right hand side (r.h.s.)
as $R^0_j$. The procedure starts by evaluating the vector
$L^0_j-R^0_j$ for all its components $j$,  in the initial density
data $\rho^0(\omega_j)$. Then, this vector becomes a function of the
Lagrange multipliers and the value of $q$. Next,  the values of
those parameters $\alpha^0_E,\alpha^0_P,\alpha^0_N$, $\gamma^0$ and
$q^0$ that minimize the quadratic difference $\sum_j
(L^0_j-R^0_j)^2$ were determined. Finalizing this first step, the
set of obtained multipliers were substituted in the expression for
$R^0_j$ together with the initial density $\rho^0(\omega_j)$,
defining in this way  $L^1_j$ as the l.h.s. of (\ref{extremo2}). The
new density data to be employed in the next iteration
$\rho^1(\omega_j)$ were determined from
$\rho^1(\omega_j)=\exp[L^1_j]/\left(\sum_k 2 \pi r_k \Delta r_k
\exp[L^1_k]\right)$, where the denominator shows that the
normalization is imposed. The new step continues by writing a new
vector $L^1_j-R^1_j$ in which the density $\rho^1(\omega_j)$ is
substituted  in $R^1_j$ and the parameters are set free again. Then,
the expression $\sum_j(L^1_j-R^1_j)^2$ is again minimized to get the
new optimal multipliers and $q$. This iterative procedure was
carried out thirty times and the resulting degree of convergence is
expressed by the value $\sum_j (L^{30}_j-R^{30}_j)^2=2.58 \times
10^{-7}$ in comparison with the initial evaluation giving a result
$\sum_j (L^{0}_j-R^{0}_j)^2= 0.33$. Another indicator of the
convergence is Eq. (\ref{integral}). In this case the l.h.s goes
from -6742 in the first step,  down to 0.1 in the last one. Thus,
the numerical  solution of the system was taken as the $30^{th}$
iterative results for the density values $\rho^{30}(\omega_j)$ and
the set of parameters $\alpha^{30}_E,\alpha^{30}_P,\alpha^{30}_N$,
$\gamma^{30}$ and $q^{30}$.

\begin{figure}[h]
\begin{flushleft}
\hspace*{-0.4cm}
\includegraphics[width=12.5cm]{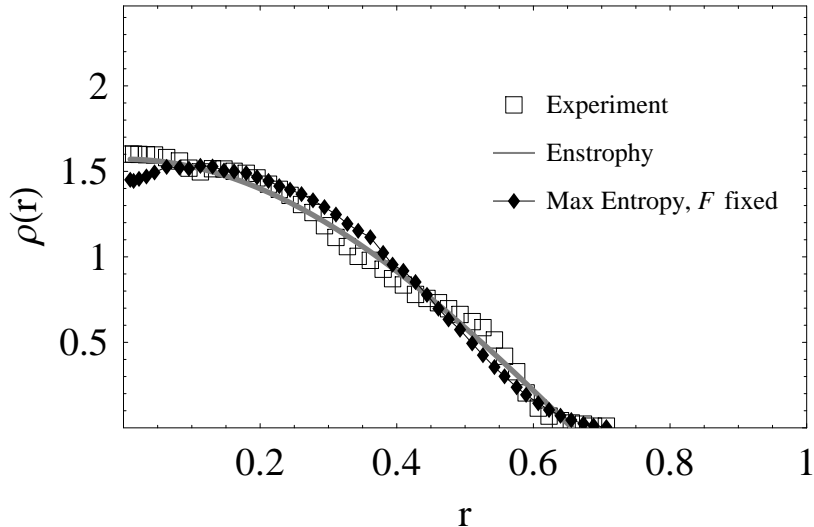}
\end{flushleft}
\caption{Density of particles $\rho(r)$ vs. the radial position. The
squares  show the experimental result $\rho^0(r_j)$ of Ref.
[\onlinecite{driscol}]. The continuous curve illustrates the density
obtained from the minimal enstrophy  model.  The diamonds indicate
the density evaluated  from the  iterative solution of the maximum
entropy principle subject to the constraint $E_q$: $\rho^{30}(r_j)$
.} \label{figura1}
\end{figure}

In Fig.\ref{figura1}  the data of Huang-Driscoll~\cite{driscol} are
depicted in comparison with the ones predicted by the restricted
minimum enstrophy model and our iterative solution. The values of
the parameters  obtained from the iterative procedure are given by
\begin{eqnarray}
a^{30}_E&=&(3.9 \pm 0.3)10^5, \\
a^{30}_N&=&(1.8 \pm 0.1)10^5, \nonumber\\
a^{30}_P&=&-4.627 \pm 0.001, \nonumber\\
\gamma^{30}&=&(3.9 \pm 0.3)10^5, \nonumber\\
q^{30}&=&0.999993 \pm 0.000001. \nonumber
\end{eqnarray}

Both for the parameters and the physical magnitudes, the errors are
given by the difference between the value of the quantity in the
last iteration $30^{th}$ and in the previous $29^{th}$. The values
obtained for the physical magnitudes are
\begin{eqnarray}
S&=&0.0716522 \pm  0.000001,\\
H&=&0.476445\pm 0.000003, \nonumber\\
P_{\theta}&=&0.866057\pm 0.0000001,\nonumber
\end{eqnarray}
which are reasonably close to the experimental values \cite{driscol}
given the natural presence of experimental errors and our discrete
approximation of the integrals. The experimental values are
$S^{exp}=0.087$ and $P^{exp}_{\theta}=0.861$.

\begin{figure}[h]
\begin{flushleft}
\hspace*{-0.4cm}
\includegraphics[width=12.5cm]{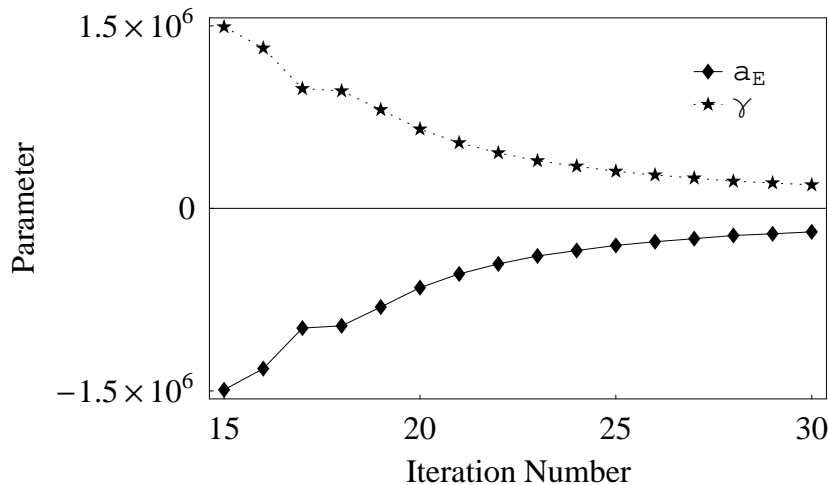}
\end{flushleft}
\caption{Lagrange multipliers $a_E$ and $\gamma$  as functions of
the iteration number. In spite of being relatively  large numbers,
their  behaviors do not exhibit instabilities in their values when
varying the step. The observed property is also fulfilled for the
multiplier of the number of particles $a_N$. } \label{multi}
\end{figure}

 It can be noted that the results for the Lagrange multipliers
 associated to the Energy constraint, the Tsallis constraint like condition
 an the number of particles requirement are relatively high
 quantities.  This  curious outcome results to be compatible with the
 resulting value of $q$ which is very close to the unit. In this case,
 the only way in which the finite differences between the Gibbs and the present result
 for the density could arise  is due to those large values
 for the multipliers. This is so because because the additional constraint for such
 a value of $q$ very close to the unit is very close to the Gibbs energy constraint.
  In order to  rule out the existence of a
 possible instability of the results for the multipliers, as a
 function of the step in the iterative process, in Fig. \ref{multi}
 we plotted the multipliers for the Energy and Tsallis constraints
 versus the iteration number. As it can be seen,
 the results continuously vary and decrease in moduli with the increase of the iterative step.

In order to search for the sensibility of the results on the way of
evaluating the integrations, the iterative method described was
carry out as well with a different way to approximate the integrals.
They were now calculated by taking a trapezoidal approach: $\int_0^R
g(r)dr\approx \sum_{i=1}^{n-1}\Delta r_i
\frac{g(r_i)+g(r_{i+1})}{2}$, with $\Delta r_k=r_{k+1}-r_k$. The
obtained results are basically the same as the ones showed here for:
the density solution of Eq.(\ref{extremo2}), the magnitudes of the
lagrange multipliers and the convergence of l.h.s from
Eq.(\ref{integral}) to zero.

The Fig. \ref{figura1}, indicates  that our solution reasonably
agrees with the experimental data presented in Ref. \cite{driscol}.
At the points with bigger values of the density the results are
close to the ones predicted by restricted enstrophy calculation.
However, comparing the curves, one can appreciate that the minimum
enstrophy model has an abrupt ending for a given radius $r_0$, which
is a limitation of this approach. In this case, the density is
written as an analytical expression given by
$\rho(r)=\alpha(J_0(\beta r)-J_0(\beta r_0))$, for $r \leq r_0$ or
vanishes otherwise. In contrast, the present discussion correctly
predicts the experimental smooth decay of the density to zero at
large distances. Thus, in that region,  the proposed procedure is
able to avoid the singularity predicted  by the minimal enstrophy
analysis. It can be concluded that the performed numerical study
shows that the statistical mechanics proposal investigated in this
work, reasonably well  describes the metaequilibrium state found in
Ref. [\onlinecite{driscol}].
\begin{figure}[h]
\begin{flushleft}
\hspace*{-0.4cm}
\includegraphics[width=12.5cm]{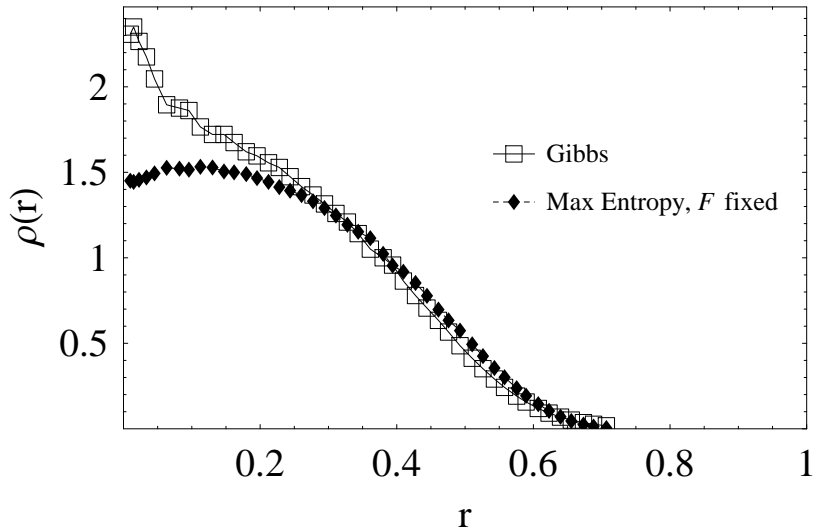}
\end{flushleft}
\caption{Comparison of the results for the density obtained by
disregarding or not the Tsallis like constraint.  The open squares
show the Gibbs results, that is the solution of the iterative
procedure by only retaining the fixed energy constraint.  The
diamonds again indicate the density evaluated from the iterative
solution of the maximum entropy principle subject to both: fixed
energy condition and the new constraint $E_q$.} \label{gibbs}
\end{figure}

Finally, in Fig. \ref{gibbs} and for comparison purposes, the
obtained density is plotted in common with the density evaluated
from the same procedure but in which the Tsallis like constraint
$E_q$ is not considered: $\gamma=0$. That is: the conserved
magnitudes are Energy, Angular\ Momentum and Normalization. Both
results drastically deviate from each other at small radial
distances. A recent work studying the behavior of a similar system
\cite{prlnew} establishes that in certain ranges of the ratio
$H/P_{\theta}$ it could dominate an evolution guided by a near
maximum Entropy principle (for bigger ratio), or alternatively, a
minimum Enstrophy (for smaller ratio). The Huang-Driscoll experiment
furnishes a primer  of the second case, since  it is approximately
well described by a minimum Enstrophy state. Therefore, as we are
able to reasonable well  describe  their data \cite{driscol}, our
results deviate from the Gibbs behavior, and furnish support to the
additional constraint argued here to be relevant in determining
 metastable and stationary states.

It should be underlined that an issue that is yet puzzling for us,
is the reason for  that very small deviations from the unit of the
$q$ number (and its associated large values of the multipliers), are
precisely the ones leading to a reasonable good match between the
numerical solution and the quasiequilibrium plasma density. We
expect to be able of considering this question elsewhere.
\section{Summary}\label{summary}

In this work,  we propose a modification of the Gibbs approach in
order to deal with stationary and metastable equilibrium states of
physical systems. As usual, the  time evolution is assumed to
maximize the value of the entropy $S$ by satisfying  the standard
constraints, like constant energy and normalization. However, the
description of the no-equilibrium states is searched by considering
that those configurations are stabilized by the presence of an extra
constraint $F$, being (approximately) conserved in the motion. The
constraint, being restricted to be time invariant, is assumed to
play the central role in stopping the evolution of the entropy to
its maximal value in the Gibbs thermal equilibrium state.
Afterwards, we assume a situation in which  the description, which
is valid for a composite system formed by two quasi independent
subsystems, is also valid for each of them. This supposition implies
that, if the constraint $ F $ commutes with density matrix, then
$F(p_{i})$ has a simplified dependence on $p_i$ of the form
$F(p_{i})=p_{i}^{q}$. One of the main outcomes of the  work arises
from this conclusion: The suggestion of an interpretation of the
Tsallis $q$ parameter as the order of the homogeneity of the
constraint $F$, when it is expressed as a function of the density
matrix. Remarkably, for small and medium radial distances the
approach furnish results of similar quality as the ones given  by
the extremum of the enstrophy and Tsallis procedures. But, moreover,
the smooth tail of the experimental density distribution at large
distances is  also predicted. Therefore, the analysis becomes able
of avoiding the non-analyticity shown by the density curve in the
mentioned alternative schemes, by describing the density decay tail
of the Huang-Driscoll experimental data  at large radial distances.
Finally, it can be referred, that the proposed description has been
also applied in Ref. [\onlinecite{carlos}],  to a gas of polaritons
contained within a quantum dot pumped by an optical resonator.
Similarly as in the here discussed plasma problem, the scheme
furnished a reasonably good description of the statistical
distributions associated to that  system.

\bigskip

\begin{acknowledgments}
A.C. and S.C. are  grateful to the Pontificia Universidad Cat\'olica
de Chile (Santiago, Chile) by the support to this research in its
initial stage. They are  also grateful to Dr. L. A. Delpino by its
collaboration and helpful discussions. Three of the authors (A.C.,
A.G. and N.G.C.) will like to deeply acknowledge the additional
support received from the Proyecto Nacional de Ciencias B\'asics
(PNCB, CITMA, Cuba) and from the Network N-35 of the Office of
External Activities (OEA) of the ICTP (Italy).  C.V. wish to
acknowledge the support received from the Department of Physics of
University of Antioquia, Medellin, Colombia.  N.G.C. thanks the
support of the BCGS of Physics and Astronomy and the Physics
Institute of Bonn University. A.C. would like to thanks Prof. Mateo
Marsili by helpful comments received while visiting the ASICTP.
Finally, S.C. would like to thank partial financial support from
FONDECYT, grant 1051075.

\end{acknowledgments}

\end{document}